 \documentclass[twocolumn, superscriptaddress, prb, showpacs]{revtex4}
 \newif\ifGALLEYversion\GALLEYversionfalse
\usepackage{ifthen}
\usepackage{graphicx}
\usepackage{amsmath}

\ifthenelse{\boolean{GALLEYversion}}{
    \marginparwidth 2.7in
    \marginparsep 0.5in
    \def\ttm#1{\marginpar{\small TT: #1}}
    \def\abm#1{\marginpar{\small AB: #1}}
    \def\bkm#1{\marginpar{\small BK: #1}}
    \def\ram#1{\marginpar{\small RA: #1}}
    \def\dpm#1{\marginpar{\small DP: #1}}
    \def\akm#1{\marginpar{\small AK: #1}}
}{
    \def\ttm#1{\relax}
    \def\abm#1{\relax}
    \def\bkm#1{\relax}
    \def\ram#1{\relax}
    \def\dpm#1{\relax}
    \def\akm#1{\relax}
}

\begin{document}

\title{Van der Waals interactions in the ground state of Mg(BH$_4$)$_2$ from density functional theory}

\author{A. Bil}
\affiliation{Department of Materials, University of Oxford, Parks Road, Oxford OX1 3PH, United Kingdom}

\author{B. Kolb}
\affiliation{Wake Forest University, Department of Physics, Winston-Salem, NC 27109 USA}

\author{R. Atkinson}
\affiliation{Wake Forest University, Department of Physics, Winston-Salem, NC 27109 USA}

\author{D. G. Pettifor}
\affiliation{Department of Materials, University of Oxford, Parks Road, Oxford OX1 3PH, United Kingdom}

\author{T. Thonhauser}
\email[E-mail: ]{thonhauser@wfu.edu}
\affiliation{Wake Forest University, Department of Physics, Winston-Salem, NC 27109 USA}

\author{A. N. Kolmogorov}
\email[E-mail: ]{aleksey.kolmogorov@materials.ox.ac.uk}
\affiliation{Department of Materials, University of Oxford, Parks Road, Oxford OX1 3PH, United Kingdom}

\date{\today}

\begin{abstract}
  { In order to resolve an outstanding discrepancy between experiment
  and theory regarding the ground-state structure of Mg(BH$_4$)$_2$,
  we examine the importance of long-range dispersive interactions on
  the compound's thermodynamic stability. Careful treatment of the
  correlation effects within a recently developed nonlocal van der
  Waals density functional (vdW-DF) leads to a good agreement with
  experiment, favoring the $\alpha$-Mg(BH$_4$)$_2$ phase (P6$_1$22)
  and a closely related Mn(BH$_4$)$_2$-prototype phase (P3$_1$12) over
  a large set of polymorphs at low temperatures. Our study
  demonstrates the need to go beyond (semi)local density functional
  approximations for a reliable description of crystalline high-valent
  metal borohydrides.}
\end{abstract}

%%%%%%%%%%%%%%%%%%%%%%%%%%%%%%%%%%%%%%%%%%%%%%%%%%%%%%%%%%%%%%%%%%%%%%%%
% 77.80.-e      Ferroelectricity and antiferroelectricity
% 71.20.-b      Electron density of states and band structure of crystalline solids
% 63.20.-e      Phonons in crystal lattices
% 78.30.-j      Infrared and Raman spectra
%%%%%%%%%%%%%%%%%%%%%%%%%%%%%%%%%%%%%%%%%%%%%%%%%%%%%%%%%%%%%%%%%%%%%%%%

\maketitle

%%%%%%%%%%%%%%%%%%%%%%%%%%%%%%%%%%%%%%%%%%%%%%%%%%%%%%%%%%%%%%%%%%%%%%%%
\section{Introduction}
\label{section.introduction}
%%%%%%%%%%%%%%%%%%%%%%%%%%%%%%%%%%%%%%%%%%%%%%%%%%%%%%%%%%%%%%%%%%%%%%%%

Light-weight metal borohydrides remain in the spotlight of
hydrogen-storage research due to their high gravimetric hydrogen
content and abundance of the constituent elements.
\cite{inorganic_rev,REVX} Li, Mg, and Ca tetraborohydrides
($M$(BH$_4$)$_n$) hold over 12 wt\% of H, but their excessive
thermodynamic stability presently renders these materials unsuitable
for practical reversible hydrogen-storage solutions.
\cite{inorganic_rev} In an effort to adjust the formation enthalpies
and the decomposition temperatures, several groups have suggested
destabilization routes via reactions with other hydrides (see e.g.\
Refs. \cite{reac_1,reac_2,reac_3}) or synthesis of mixed metal
tetraborohydrides (e.g.\ Li-Cu) \cite{mixed_Li_Cu}. Detailed knowledge
of materials' ground states becomes essential, as the possibility of
new compound formation depends on small free energy differences for
the phases involved.

Metal tetraborohydrides are held together primarily by a combination
of covalent (B-H) and ionic ([$M$]$^{n\delta +}$-[BH$_4$]$^{\delta
-}$) interactions. Packing of the fairly rigid [BH$_4$]$^{\delta -}$
units and [$M$]$^{n\delta +}$ ions into crystalline structures depends
strongly on the valency and size of the metal atom, with exceptionally
complex configurations occurring for the medium-sized divalent
magnesium. Based on the experimental data, the ground states below and
above $T=453$ K, namely $\alpha$ and $\beta$, have been assigned space
groups P6$_1$22 (330 atoms/u.c.) \cite{Mg_Dai,exp_chem_matter} and
Fddd (704 atoms/u.c.) \cite{exp_acta}, respectively. Strikingly,
numerous density functional theory (DFT) studies
\cite{Mg_Nakamori,Mg_Vajeeston,Mg_Dai,Mg_Voss,Mg_Ozolins,
Mg_Zhou,Mg_Caputo} have converged on a completely unrelated F222
structure as the most stable low-$T$ polymorph.
\cite{Mg_Ozolins,Mg_Zhou,Mg_Caputo} Recently, it has been pointed out
\cite{Lodziana} that the lack of a proper description of dispersion
forces in the standard DFT approach may be a factor favoring
low-density F222 structure over the experimental $\alpha$ phase.

In this study we take a key step towards resolving the existing
discrepancy: we demonstrate that the weak dispersive interactions,
important in other ionic (KCl and KBr) \cite{KCl_KBr} and
covalent-ionic (Mg(OH)$_2$ and Ca(OH)$_2$) \cite{B3LYP-D*-2} systems,
indeed play a critical role in defining the Mg(BH$_4$)$_2$ ground state.
We show that the commonly used Ceperley-Alder \cite{LDA} or Perdew,
Burke, and Ernzerhof (PBE) \cite{PBE} functionals artificially
stabilize configurations with unusually high and low density,
respectively. Inclusion of the dispersive contributions via a nonlocal
van der Waals density functional (vdW-DF)
\cite{Langreth_1,Thonhauser1} or as a semi-empirical PBE-D* correction
\cite{Grimme1,Grimme2,B3LYP-D*-1} changes the relative stability of
considered structures and favors the layered motif \cite{Mn} occurring
in related $\alpha$-Mg(BH$_4$)$_2$ and Mn(BH$_4$)$_2$ prototypes.

The careful re-examination of Mg(BH$_4$)$_2$ has become possible due
to a recent surge of attempts to include vdW interactions in the DFT.
\cite{Grimme_rev,Langreth_rev} The truly nonlocal correlation
functional in the vdW-DF approach has shown good transferability for a
range of systems reaching from simple dimers \cite{dimers} and
physisorbed molecules \cite{physisorption} to DNA \cite{DNA} and drug
design \cite{drug}. An efficient FFT formulation of vdW-DF
\cite{Soler} has allowed us to calculate the $T=0$ K relative
stability for Mg(BH$_4$)$_2$ structures of unprecedented size (up to
330 atom/u.c.). Calculation of the Gibbs energy vibrational
contributions at finite temperatures is much more computationally
demanding. In order to examine the relative stability as a function of
temperature we have employed a semi-empirical method developed by
Grimme \cite{Grimme1,Grimme2}, that includes the long-range
contributions via damped pairwise $f_{\text{dmp}}(R)C_6R^{-6}$ terms
at a negligible cost compared to standard DFT calculations. The method
has gained popularity providing an improved description of molecular
systems. However, Civalleri {\it et al.} \cite{B3LYP-D*-1,B3LYP-D*-2}
demonstrated the need to adjust the parameterization (from PBE-D to
PBE-D*) for the application of the method to crystalline solids and we
observe a better transferrability of the modified set for the
challenging case of Mg(BH$_4$)$_2$.

We describe the considered library of structure types and the
simulation settings in Sec. \ref{section.settings}, present our
systematic comparison of the performance of five DFT-based methods for
Mg(BH$_4$)$_2$ at $T=0$ K in Sec. \ref{section.comparison},
demonstrate the effect of vibrational entropy on the polymorphs'
relative stability at finite temperatures in
Sec. \ref{section.phonons}, and conclude in
Sec. \ref{section.conclusions}.

%%%%%%%%%%%%%%%%%%%%%%%%%%%%%%%%%%%%%%%%%%%%%%%%%%%%%%%%%%%%%%%%%%%%%%%%
\section{Simulation setup}
\label{section.settings}
%%%%%%%%%%%%%%%%%%%%%%%%%%%%%%%%%%%%%%%%%%%%%%%%%%%%%%%%%%%%%%%%%%%%%%%%

{\it Library of structure types.} In addition to a large pool of
previously considered candidate structures, we include potentially
relevant A(BC$_4$)$_2$ prototypes found in the Inorganic Crystal
Structure Database (ICSD) \cite{ICSD}; the full list of 36 structures
is given in Table \ref{T1}. All energies are referenced to the
experimental low-$T$ $\alpha$ phase, which has been recently argued to
have a higher symmetry (P6$_1$22) \cite{Mg_Dai,exp_chem_matter} than
originally thought (P6$_1$) \cite{exp_acta,exp_angewandte}. Our
simulations confirm that, while the two structures are nearly
degenerate in energy, only P6$_1$22 is dynamically stable (P6$_1$ has
multiple imaginary phonon modes reaching $50i$ cm$^{-1}$). The
structure of the high-$T$ $\beta$-phase, observed experimentally to be
stable between 453 K and 613 K \cite{exp_angewandte,exp_acta}, remains
an open question: the first powder diffraction orthorhombic solution
with the Fddd symmetry, based on the positions of the B and Mg atoms,
has six imaginary phonon modes at $\Gamma$ reaching $65i$
cm$^{-1}$. Notable Mg(BH$_4$)$_2$ candidates identified in previous
DFT studies include the trigonal P$\overline{3}$m1 structures
\cite{Mg_Nakamori} as well as the unusual low-density F222 phase
derived from I$\overline{4}$m2 \cite{Mg_Ozolins,Mg_Zhou}. Figures
1a--c illustrate the diversity of the morphologies as different
arrangements of the Mg$^{2\delta +}$ ions and BH$_4^{\delta -}$ units
result in layered, hollow-framework-like, or fairly uniform
densely-packed structures.

{\it Simulation settings.} The total energy calculations are carried
out in the generalized gradient approximation (GGA) with the PBE
\cite{PBE} exchange-correlation ({\it xc}) functional and the local
density functional approximation (LDA) \cite{LDA} as implemented in
{\small VASP} \cite{kresse1993}. We employ projector augmented-wave
pseudopotentials (PAW) \cite{bloechl994} in which the semicore states
are treated as valence. An energy cutoff of 500 eV and dense
Monkhorst-Pack ${\bf k}$-point meshes ($\sim$ 0.03\AA$^{-1}$ in each
direction in the Brillouin zone) \cite{MONKHORST_PACK} are
applied. The vdW-DF, PBE, PBE-D, and PBE-D* calculations are performed
with a modified version of PWscf \cite{PWscf} using ultrasoft
pseudopotentials. All structures were fully relaxed with the threshold
of 10$^{-5}$ [Ry] for energy convergence and 3$\cdot$10$^{-1}$ [Ry
a.u.$^{-1}$] for residual force; the residual stress was typically
under 1 kbar. Benchmark tests given in Appendix A suggest that for a
given {\it xc} functional the errors arising from other factors
(convergence criteria, choice of pseudopotential, ect.) are rather
small allowing us to resolve polymorphs $\sim 0.005$ eV/BH$_4$ apart.
Phonon spectra and the Gibbs energy corrections for selected
structures are calculated with PBE and PBE-D* with a finite
displacement method as implemented in PHON \cite{PHON}. Forces for
phonon calculations were obtained from {\small VASP} (PBE) or QMPOT
\cite{QMPOT,Sauer} linked to {\small VASP} (PBE-D*) (for more details
see Sec. \ref{section.phonons}).

\textit{ Semiempirical dispersion contributions.} The Grimme
corrections\cite{Grimme1,Grimme2} are introduced as
\begin{eqnarray}
  E_{\text{disp}} = -s_6\sum_{i=1}^{N_{\text{at}}-1} \sum_{j=i+1}^{N_{\text{at}}}\frac{C_6^{ij}}{R_{ij}^6}f_{\text{dmp}}(R_{ij}),
\end{eqnarray}
where $C_6^{ij} =\sqrt{C_6^{i}C_6^{j}}$, $f_{\text{dmp}} =
(1+e^{-d(R_{ij}/R_r-1)})^{-1}$, and $R_r$ is the sum of individual
$R_0$.

For PBE-D calculations\cite{Grimme2} we used the following set of
C$_6$ coefficients [Jnm$^6$mol$^{-1}$] and van der Waals radii $R_0$
[\AA]: for H atom C$_6$ = 0.14, R$_0$ =1.001; for B atom C$_6$ = 3.13,
$R_0$ =1.485; for Mg atom C$_6$ = 5.71, $R_0$ =1.364. Dimensionless
parameter in the damping function and global scaling parameter are $d$
= 20 and $s_6$ = 0.75, respectively. For PBE-D*
parameterization\cite{B3LYP-D*-1} $R_0$ was multiplied by 1.3 (H), or
by 1.05 factor (B, Mg).
%
%%%%%%%%%%%%%%%%%%Figure 1%%%%%%%%%%%%%%%%%%%%%%%%%%%%%%%%%%%%
%
\begin{figure}[t]
\begin{center}
%\vspace{-0.5 cm}
\hspace{-0.5 cm}
\includegraphics[width=70mm,angle=0]{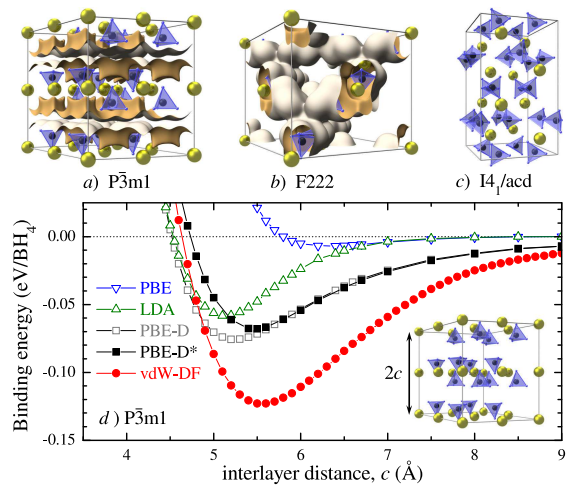}
\caption{ \small (Color online) $a$--$c$) Hypothetical structures of
  Mg(BH$_4$)$_2$ with magnesium, boron, and hydrogen atoms shown as
  large (yellow), medium (black), and small (blue) spheres. The
  isosurfaces corresponding to charge density of 0.06 e\AA$^{-3}$
  illustrate the layered and the 3D hollow framework structure of
  P$\overline{3}$m1 and F222 phases, respectively; anisotropy of the
  charge density in I4$_1$/acd (not shown) is less pronounced. $d$)
  Binding energy as a function of the interlayer spacing in
  P$\overline{3}$m1-Mg(BH$_4$)$_2$; the layers are allowed to relax
  but the intralayer distortions are insignificant.}
\label{F1}\end{center}
\end{figure}
%
%%%%%%%%%%%%%%%%%%Figure 1%%%%%%%%%%%%%%%%%%%%%%%%%%%%%%%%%%%%

%%%%%%%%%%%%%%%%%%%%%%%%%%%%%%%%%%%%%%%%%%%%%%%%%%%%%%%%%%%%%%%%%%%%%%%%
\section{Performance of DFT-based methods for Mg(BH$_4$)$_2$}
\label{section.comparison}
%%%%%%%%%%%%%%%%%%%%%%%%%%%%%%%%%%%%%%%%%%%%%%%%%%%%%%%%%%%%%%%%%%%%%%%%

%%%%%%%%%%%%%%%%%%Figure 2%%%%%%%%%%%%%%%%%%%%%%%%%%%%%%%%%%%%
\begin{figure*}[t!]
%\begin{center}
%\vspace{-0.5 cm}
%\hspace{-0.8 cm}
\includegraphics[width=170mm,angle=0]{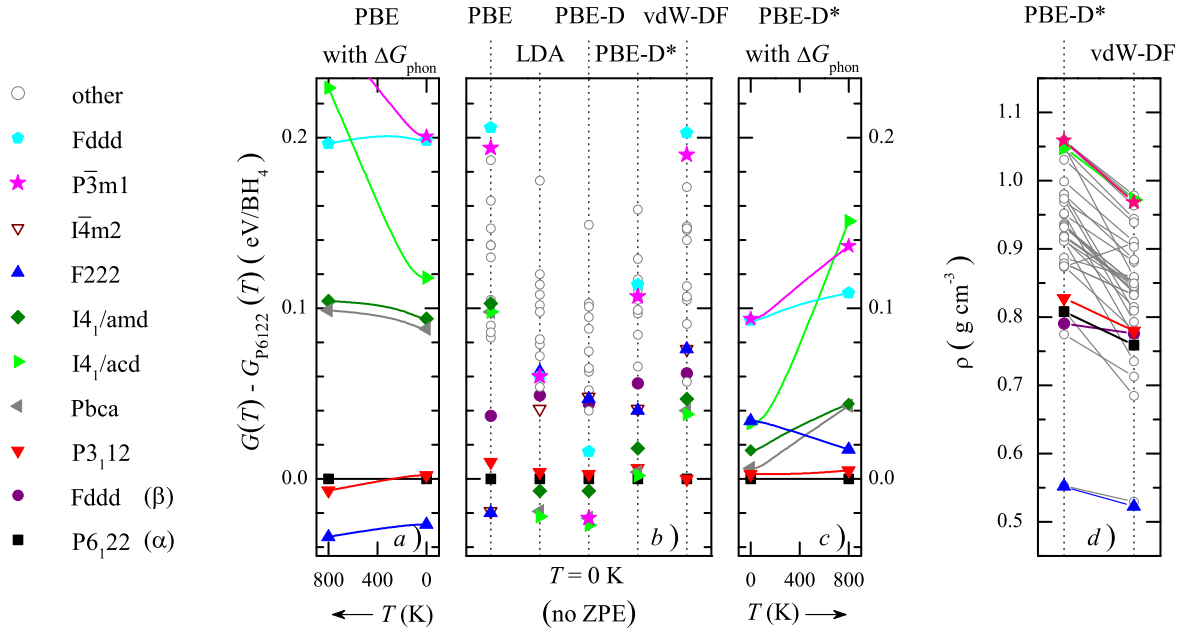}
\caption{ \small (Color online) $a$--$c$) Relative Gibbs energy of
  selected Mg(BH$_4$)$_2$ polymorphs referenced to the P6$_1$22 phase:
  in $b$) the relative energy is shown for all five approximations at
  $T$=0 K without zero point energy while in $a$,$c$) vibrational
  contributions to $G(T)$ are included for PBE and PBE-D*. $d$)
  Density of Mg(BH$_4$)$_2$ polymorphs fully relaxed with PBE-D*
  and vdW-DF methods ($\rho_{\alpha}^{exp}=0.782$ g cm$^{-3}$)
  \cite{exp_chem_matter}.}
\label{F2}
%\end{center}
\end{figure*}
%%%%%%%%%%%%%%%%%%Figure 2%%%%%%%%%%%%%%%%%%%%%%%%%%%%%%%%%%%%
%
The necessity to treat Mg(BH$_4$)$_2$ beyond the standard DFT
approximations becomes evident when the PBE functional shows
virtually no interlayer binding for the exemplary layered
P$\overline{3}$m1 phase (Fig.\ 1d). The lack of a noticeable
covalent or electrostatic interaction between the layers is a
result of the particular packing of the BH$_4$ tetrahedra that
makes the interlayer interface consist of two parallel sheets of
hydrogen atoms already engaged in covalent B-H bonds. Introduction
of the vdW interactions via vdW-DF, PBE-D, and PBE-D* leads to
interlayer cohesions of 0.123, 0.076, and 0.068 eV/BH$_4$,
respectively. The size of the extra binding is substantial
considering that the reported energy difference between the
experimental ($\alpha$) and theoretical (F222) ground states
within PBE is only 0.024 eV/BH$_4$ \cite{Mg_Caputo}. The LDA,
known to mimic dispersive interactions (e.g., in graphite
\cite{ak06}), also gives a 0.058 eV/BH$_4$ binding, but -- as we
show below -- should not be used as a substitute for a properly
constructed nonlocal functional.

We compare the performance of the five different methods, PBE, LDA,
PBE-D, PBE-D*, and vdW-DF, by plotting the relative stability and
compound density for selected structures in Fig.\ 2 (the corresponding
values for all 36 structures are listed in Table I).  A number of
unexpected features emerge as we examine the total energy data moving
from left to right in Fig.\ 2b. At the PBE level, we find the highest
stability of the least compact F222 phase, in agreement with
Ref. \cite{Mg_Voss}. However, in the LDA, arguably better suited for
simulating Mg(BH$_4$)$_2$ based on our test in Fig.\ 1, there is a
dramatic change in the ordering of the polymorphs' enthalpies with the
ground state now being the most compact I4$_1$/acd structure. It is
evident that, compared to PBE, the LDA favors higher packing: it
lowers the relative enthalpy of the two densest P$\overline{3}$m1 and
I4$_1$/acd polymorphs by 0.134 and 0.120 eV/BH$_4$, respectively, and
increases the relative enthalpy of the sparsest F222 by 0.083
eV/BH$_4$.
%
%%%%%%%%%%%%%%%%%%Figure 3%%%%%%%%%%%%%%%%%%%%%%%%%%%%%%%%%%%%
%
\begin{figure}[t!]
\begin{center}
%\vspace{-0.5 cm}
\hspace{-0.5 cm}
\includegraphics[width=88mm,angle=0]{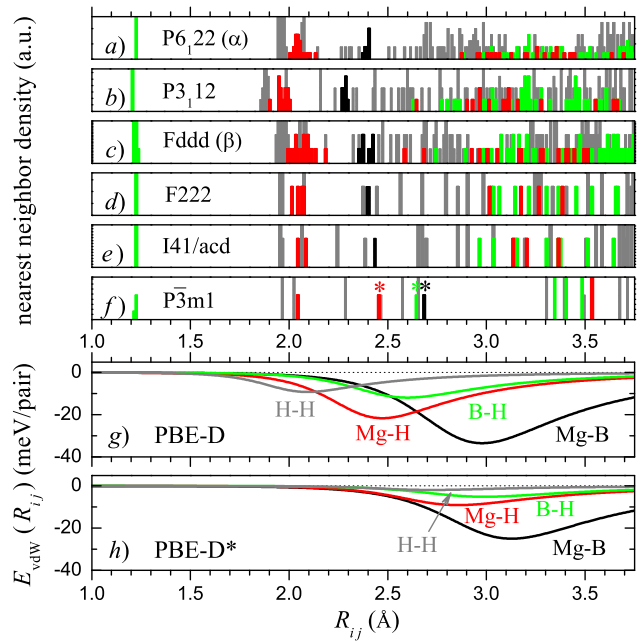}
\caption{ \small (Color online) $a-f)$ Nearest neighbor histograms in
  selected structures fully relaxed in PBE-D*; the stars in $f$) mark
  intralayer distances.  $g,h)$ Strength of pairwise Grimme
  corrections in the PBE-D and PBE-D* parameterizations.}
\label{F3}\end{center}
\end{figure}
%
%
%%%%%%%%%%%%%%%%%%Figure 3%%%%%%%%%%%%%%%%%%%%%%%%%%%%%%%%%%%%
%
%%%%%%%%%%%%%%%%%%Figure 4%%%%%%%%%%%%%%%%%%%%%%%%%%%%%%%%%%%%
%
\begin{figure}[t!]
\begin{center}
%\vspace{-0.5 cm}
%\hspace{-0.5 cm}
\includegraphics[width=88mm,angle=0]{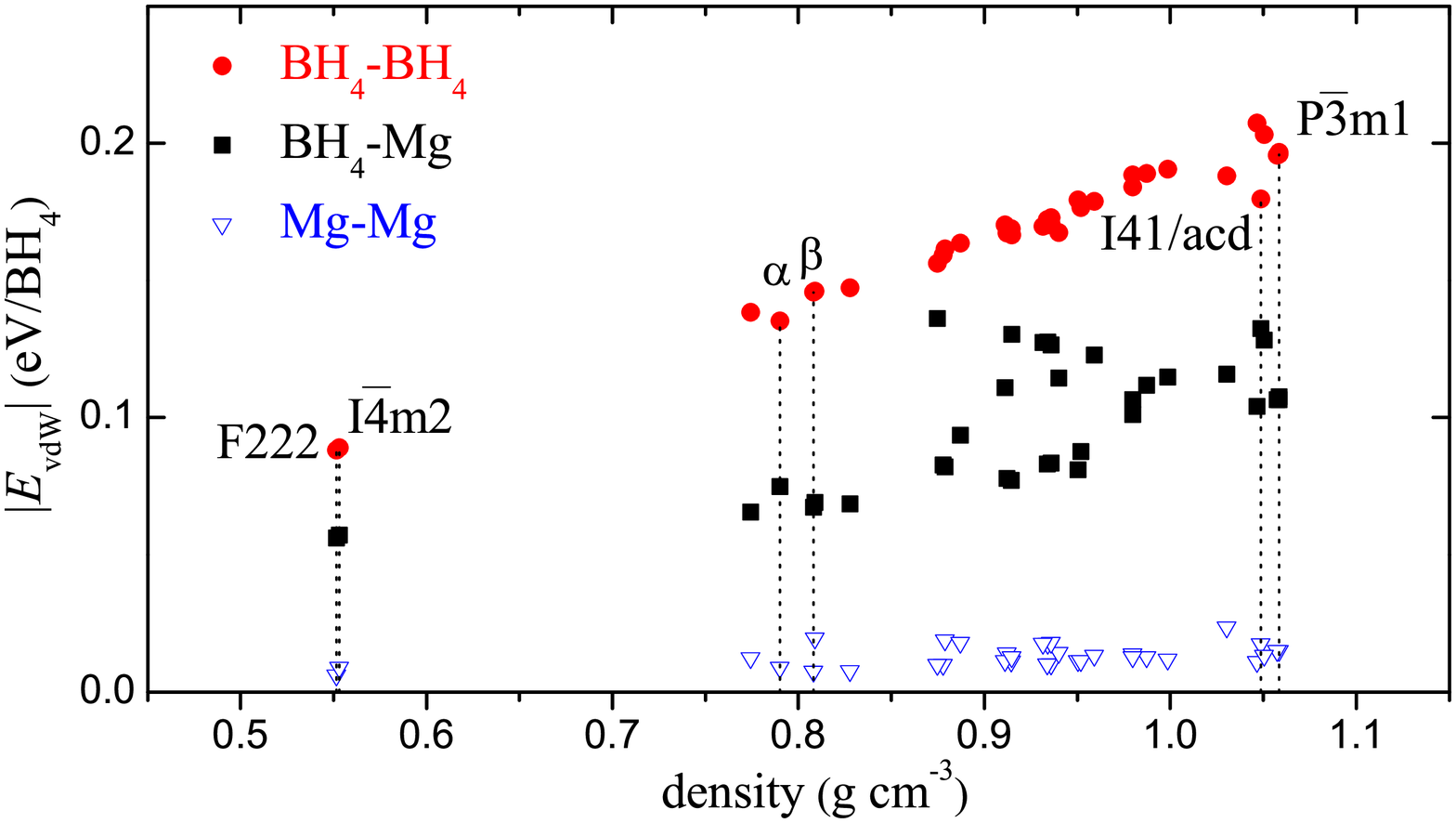}
\caption{ \small (Color online) Breakdown of dispersion energy
contributions calculated at PBE-D* level as a function of
polymorph density for Mg(BH$_4$)$_2$ structures listed in
Table \ref{T1}. } \label{F4}\end{center}
\end{figure}
%%%%%%%%%%%%%%%%%%Figure 3%%%%%%%%%%%%%%%%%%%%%%%%%%%%%%%%%%%%

Inclusion of the dispersive contribution in PBE-D also leads to a
significant reordering of the relative stabilities obtained with
PBE and promotes the I4$_1$/acd structure. In fact, there is more
agreement between the LDA vs PBE-D rather than the PBE vs PBE-D
sets. The P$\overline{3}$m1 polymorph gets an even larger 0.217
eV/BH$_4$ gain in relative stability dropping below $\alpha$ by
0.023 eV/BH$_4$, which may look surprising given the similar size
of the interlayer binding for this structure in LDA and PBE-D
(Fig.\ 1). To help explain this result we plot pairwise
interaction strengths (Fig.\ 3g) and histograms of nearest
neighbors (NNs) for relevant structures (Fig.\ 3a--f). It has been
discussed \cite{B3LYP-D*-1,Grimme_rev} that introduction of the
dispersive interactions `on top' of DFT requires a smooth cut-off
of the attractive $C_{ij}/R_{ij}^{-6}$ terms just below the
typical vdW distances to avoid their unphysical contributions at
short $R_{ij}$. Panels $f$ and $g$ in Fig.\ 3 show that a number
of intralayer distances in P$\overline{3}$m1 (marked with stars)
happen to fall in the critical range, around the corresponding
$R_{i}^{vdW}+R_{j}^{vdW}$ values fitted to molecular datasets,
which can give rise to artificially strong binding.

The suggested adjustment of the vdW parameters within PBE-D*
\cite{B3LYP-D*-1} helps reduce the vdW overbinding for the dense
P$\overline{3}$m1 and I4$_1$/acd structures and for the first time we
observe the $\alpha$-phase to be the most stable polymorph among the
considered diverse set of candidates, albeit by a small margin. The
classical pairwise representation of the vdW corrections in the Grimme
approach (Eq. 1) makes it possible to examine the individual vdW
contributions. According to Fig.\ 3h, the increase in $R_{i}^{vdW}$
values (from 1.001, 1.485, 1.364 \AA\ to 1.301, 1.559, 1.432 \AA\ for
$i$ = H, B, Mg, respectively) greatly diminishes the direct H-H
interaction. However, Fig.\ 4 shows that the BH$_4$-BH$_4$ interaction
still dominates the long-range binding, which -- not unexpectedly --
scales almost linearly with the crystal density. The second largest
contribution, Mg-BH$_4$, deviates from linear dependence due to a
variable coordination of Mg ions. The classical attractive terms cause
up to $\sim 20\%$ volume reduction compared to structures optimized
with PBE.

Finally, the vdW-DF method correctly predicts $\alpha$ to be the
low-$T$ ground state and separates it from all other unrelated
structures by a considerable 0.040 eV/BH$_4$; the near degeneracy
of $\alpha$ and P3$_1$12 (Mn(BH$_4$)$_2$ prototype) can be traced
to the close structural relationship between the two layered
structures discussed in detail in Ref. \cite{Mn}. The method
places the hypothetical F222 structure 0.076 eV/BH$_4$ above
$\alpha$ and differentiates between the two high-density 2D
P$\overline{3}$m1 and 3D I4$_1$/acd structures by improving the
stability of just the latter. This suggests that the
overestimation of the intralayer binding in the difficult
P$\overline{3}$m1 case might still not be fully corrected within
PBE-D*. Comparison of crystal densities for PBE-D* and vdW-DF in
Fig.\ 2d shows that the latter results in less compact structures;
the effect of the nonlocal correlations on the volume change is
not easy to quantify because the method is based on the exchange
and correlation functionals taken from revPBE \cite{revPBE} and
LDA, respectively. The calculated vdW-DF value of $\rho=0.758$ g$
$cm$^{-3}$ for the $\alpha$ phase is 3\% below the experimental
value which is consistent with a typical 1\% overestimation of
bond lengths for GGA-based methods.

%%%%%%%%%%%%%%%%%%%%%%%%%%%%%%%%%%%%%%%%%%%%%%%%%%%%%%%%%%%%%%%%%%%%%%%%
\section{Inclusion of the vibrational contributions to $G(T)$}
\label{section.phonons}
%%%%%%%%%%%%%%%%%%%%%%%%%%%%%%%%%%%%%%%%%%%%%%%%%%%%%%%%%%%%%%%%%%%%%%%

Inclusion of finite temperature contributions due to the vibrational
entropy is orders of magnitude more expensive than the calculation of
the total energy at zero temperature. We estimate the vibrational
corrections using two DFT-based methods, PBE-D and PBE-D*, which is
computationally feasible and allows us to show directly the effect of
the vdW interactions. Hessian matrices were calculated in a numerical
fashion from analitycal forces and atomic displacements for
sufficiently large supercells (at least 88 atoms).\cite{supercells}
According to our extensive tests for Mg(BH$_4$)$_2$, the 0.00133 \AA\
displacements and the 10$^{-8}$ eV SCF energy convergence criterion
ensured minimal errors ($\sim$ 10 cm$^{-1}$) for the calculated
frequencies coming from the unharmonic effects and numerical
factors. $q$-mesh for the dynamical matrices was typically two times
denser than the $k$-mesh for the energy calculations in each
direction. The analitycal forces at the PBE level were found with
{\small VASP} for fully relaxed polymorph structures. The analytical
forces at the PBE-D* level were calculated using {\small QMPOT} linked
with {\small VASP}. In this case, since {\small QMPOT} does not
currently offer stress calculations, the structures were optimized as
follows: the unit cells were first fully relaxed with PWscf and then
the ionic positions were re-optimized with a combination of {\small
QMPOT} and {\small VASP}. As described in Appendix A, our careful
tests for selected structures showed insignificant variation in the
equilibrium cell parameters in the two settings. Once the phonon
densities of states $n(\omega)$ were found, the vibrational
contributions $G_{phon}(T)$ were included via \cite{PHON}
\begin{eqnarray}
  G_{phon}(T) = k_{\text{B}}T \int_{0}^{\infty} \ln \left[ 2\sinh(\hbar\omega/(2 k_{\text{B}}T) \right] n(\omega) d \omega
\end{eqnarray}
%
%%%%%%%%%%%%%%%%%%%%%%%%%%%%%%%Table 1%%%%%%%%%%%%%%%%%%%%%%%%%%%%%%%%%%%%%%%%%%%%%%%%%%%%
%
%\vspace{-5mm}
\begin{widetext}
\begin{center}
\begin{table}[!b]
\begin{tabular}{c|c|c|c|c|c|c|c|c|c|c|c|c|c|c}
\hline
  &                    &                          &     &   \multicolumn{2}{c|}{vDW-DF} & \multicolumn{2}{c|}{ PBE-D*} & \multicolumn{2}{c|}{PBE-D}& \multicolumn{2}{c|}{PBE} & \multicolumn{2}{c|}{LDA}   \\
\hline
Nr&Parent str.         & Symm.                    & $Z$ &$\varrho$ &$\Delta E$ &$\varrho$          &$\Delta E$ &$\varrho$  &$\Delta E$ &$\varrho$  &$\Delta E$ &$\varrho$  &$\Delta E$   \\
\hline
36&Mg(AlH$_4$)$_2$     &  P$\overline{3}m1$ (164) &  1  & 0.780      &0.439    &  0.875              &0.550    & 0.948       &0.543    & 0.735       &0.606    & 0.975       &0.684  \\
35&Mg(BH$_4$)$_2    $  &  P2/m (10)                    &  2  & 0.793      &0.405    &  0.915              &0.447    & 1.061       &0.345    & 0.807       &0.517    & 1.112       &0.433   \\
34&Mg(BH$_4$)$_2$      & P$\overline{3}$  (147)   &  9  & 0.845      &0.341    &  0.936              &0.322    & 0.985       &0.241    & 0.848       &0.409    & 1.037       &0.350   \\
33&Zn(ReO$_4$)$_2$     & P$\overline{3}$  (147)   &  1  & 0.850      &0.340    &  0.932              &0.319    & 0.981       &0.238    & 0.850       &0.410    & 1.040       &0.350   \\
32&ZrMo$_2$O$_8$       & P$\overline{3}1c$ (163)  &  6  & 0.848      &0.340    &  0.934              &0.319    & 0.979       &0.238    & 0.850       &0.409    & 1.028       &0.350   \\
31&Mn(BF$_4$)$_2$      & Pnma (62)                &  4  & 0.847      &0.278    &  0.940              &0.251    & 0.993       &0.175    & 0.852       &0.318    & 1.046       &0.258   \\
30&Mg(BH$_4$)$_2$      & Pnnm (58)                &  2  & 0.735      &0.266    &  0.911              &0.270    & 0.980       &0.250    & 0.775       &0.320    & 1.015       &0.297   \\
29&Cr(AlCl$_4$)$_2$    & Pca2$_1$ (29)            &  4  & 0.809      &0.248    &  0.980              &0.193    & 1.051       &0.148    & 0.787       &0.090    & 1.085       &0.189   \\
28&Mg(BH$_4$)$_2$      & P2$_1$/c (14)            &  2  & 0.872      &0.245    &  0.959              &0.173    & 1.011       &0.092    & 0.816       &0.186    & 1.052       &0.174   \\
27&Mg(BH$_4$)$_2$      & P2$_1$ (4)               &  2  & 0.859      &0.212    &  1.050              &0.167    & 1.117       &0.091    & 0.846       &0.210    & 1.149       &0.157   \\
26&Ca(BH$_4$)$_2$      & Fddd (70)                &  8  & 0.903      &0.203    &  0.987              &0.114    & 1.043       &0.016    & 0.943       &0.206    & 1.090       &0.093   \\
25&Mg(BH$_4$)$_2$      & P2/m (10)                &  2  & 0.963      &0.195    &  1.058              &0.106    & 1.101       &-0.023   & 0.946       &0.193    & 1.157       &0.060   \\
24&Sr(AlCl$_4$)$_2$    & P2/c (13)              &  2  & 0.978      &0.194    &  1.058              &0.107    & 1.101       &-0.023   & 0.947       &0.194    & 1.158       &0.060   \\
23&Mg(BH$_4$)$_2$      & P$\overline{3}$1m (162)  &  6  & 0.974      &0.194    &  1.058              &0.107    & 1.101       &-0.023   & 0.947       &0.194    & 1.153       &0.060   \\
22&Mg(BH$_4$)$_2$      & P$\overline{3}$m1 (164)  &  1  & 0.968      &0.191    &  1.058              &0.107    & 1.101       &-0.023   & 0.950       &0.194    & 1.145       &0.060   \\
21&ZrMo$_2$O$_8$       & Pmn2$_1$ (31)            &  2  & 0.910      &0.182    &  0.999              &0.138    & 1.051       &0.080    & 0.894       &0.224    & 1.076       &0.143   \\
20&Be(BH$_4$)$_2$      & I4$_1$cd (110)           & 16  & 0.909      &0.173    &  0.878              &0.110    & 0.911       &0.087    & 0.677       &0.026    & 0.961       &0.093   \\
19&Mg(BH$_4$)$_2$      & P$\overline{1}$ (2)      &  4  & 0.685      &0.172    &  0.809              &0.158    & 0.896       &0.149    & 0.651       &0.147    & 0.924       &0.175   \\
18&Mg(BH$_4$)$_2$      & P1 (1)                   &  2  & 0.780      &0.149    &  0.915              &0.111    & 0.964       &0.040    & 0.760       &0.086    & 0.986       &0.098   \\
17&Ba(BF$_4$)$_2$      & P2$_1$/c (14)             &  8  & 0.884      &0.147    &  0.980              &0.117    & 1.020       &0.065    & 0.874       &0.187    & 1.076       &0.108   \\
16&Be(AlO$_4$)$_2$     & P$\overline{1}$ (2)      &  2  & 0.770      &0.147    &  0.952              &0.129    & 0.984       &0.103    & 0.823       &0.163    & 1.031       &0.082   \\
15&Mg(BH$_4$)$_2$      & P2/c   (13)              &  4  & 0.857      &0.140    &  0.887              &0.113    & 0.936       &0.101    & 0.755       &0.137    & 0.985       &0.120   \\
14&Mg(BH$_4$)$_2$      & C2/c   (15)              &  4  & 0.832      &0.113    &  0.879              &0.105    & 0.923       &0.095    & 0.757       &0.130    & 0.970       &0.114   \\
13&Mg(BH$_4$)$_2$      & Pm (6)                   &  1  & 0.877      &0.107    &  0.950              &0.099    & 0.978       &0.075    & 0.841       &0.137    & 1.017       &0.079   \\
12&Pt(SO$_4$)$_2$      & P2$_1$/c (14)             &  4  & 0.713      &0.106    &  0.774              &0.097    & 0.803       &0.088    & 0.679       &0.083    & 0.844       &0.102   \\
11&Ca(BF$_4$)$_2$      & Pbca (61)                &  8  & 0.841      &0.091    &  0.912              &0.085    & 0.958       &0.063    & 0.766       &0.105    & 1.008       &0.072  \\
10&Mg(BH$_4$)$_2$      & I$\overline{4}$m2 (119)  &  4  & 0.529      &0.077    &  0.553              &0.041    & 0.553       &0.048    & 0.533       &-0.019   & 0.585       &0.041   \\
 9&Mg(BH$_4$)$_2$      & F222 (22)                &  8  & 0.522      &0.076    &  0.552              &0.040    & 0.552       &0.047    & 0.532       &-0.020   & 0.591       &0.063   \\
 8&Cd(AlCl$_4$)$_2$    & Pc (7)                 &  2  & 0.836      &0.062    &  0.936              &0.066    & 0.967       &0.052    & 0.801       &0.092    & 0.998       &0.054   \\
 $\beta$&Mg(BH$_4$)$_2$&  Fddd $(70)$             &  64 & 0.775      &0.062    &  0.790              &0.056    & 0.819       &0.044    & 0.727       &0.037    & 0.865       &0.049   \\
 6&Mg(BH$_4$)$_2$      & Pmc2$_1$ (26)            &  2  & 0.835      &0.058    &  0.934              &0.066    & 0.966       &0.052    & 0.790       &0.090    & 1.002       &0.054   \\
 5&Mg(BH$_4$)$_2$      & I4$_1$/amd (141)         &  8  & 0.939      &0.047    &  1.030              &0.018    & 1.044       &-0.007   & 0.974       &0.103    & 1.082       &-0.007  \\
 4&Sr(AlCl$_4$)$_2$    & I4$_1$/acd (142)         &  4  & 0.972      &0.043    &  1.049              &0.002    & 1.078       &-0.027   & 0.986       &0.098    & 1.102       &-0.022  \\
 3&Sr(AlCl$_4$)$_2$    & Pbca (61)                &  8  & 0.944      &0.040    &  1.047              &0.004    & 1.074       &-0.023   & 0.979       &0.098    & 1.100       &-0.019  \\
 2&Mn(BH$_4$)$_2$      & P3$_1$12 (151)           &  9  & 0.780      &0.000    &  0.828              &0.007    & 0.837       &0.003    & 0.791       &0.010    & 0.878       &0.004   \\
 $\alpha$&Mg(BH$_4$)$_2$&    P6$_1$22 $(178)$     & 30  & 0.758      &0        &  0.808              &0        & 0.814       &0         & 0.770       &0        & 0.860       &0 \\
\hline
\end{tabular}
\caption{ \small Relative stability $\Delta E$ [eV/BH$_4$] and
density $\varrho$ [g cm$^{-1}$] ($T=0$ K, no zero point energy
corrections) of magnesium borohydride polymorphs referenced to the
experimentally observed $\alpha$ phase. $Z$ stands for the number
of Mg(BH$_4$)$_2$ formulas per unit cell. In the third column
space groups information (international symbols and numbers) is
reported. }\label{T1}
\end{table}
%%%%%%%%%%%%%%%%%%%%%%%%%%%%%%%%%%%%%%%%%%%%%%%%%%%%%%%%%%%%%%%%%%%%%%%%%%%%%%%%%%%%%%%%%%%%%
\end{center}
\end{widetext}
%%%%%%%%%%%%%%%%%%%%%%%%%%%%%%%%%%%%%%%%%%%%%%%%%%%%%%%%%%%%%%%%%%%%%%%%

Calculated relative Gibbs energies $G(T)$ for selected low-energy
structures are shown in Fig.\ 2a and 2c. We find that the zero point
energy alone can change the relative stability by over 0.04
eV/BH$_4$. Addition of dispersive interactions leads to about 0.01
eV/BH$_4$ changes in the relative stability at $T=800$ K and can
affect the ordering of the polymorphs as happens for the P3$_1$12
structure. The related $\alpha$ and P3$_1$12 phases remain nearly
degenerate in the whole $T$ range and cannot be unambiguously resolved
within our vdW-DF or PBE-D* simulations. The search for the high-$T$
ground state may be simplified by our observation of little variation
in the enthalpy difference between $\alpha$ and $\beta$ for all the
five methods in Fig.\ 2: due to the close structural relationship
between the two phases, shown in Fig.\ 3a and 3c, there must be a
cancellation of errors. Hence, despite the remaining noticeable
difference between the PBE-D* and vdW-DF relative stability for some
polymorphs the former could be an appropriate choice for
identification of viable high-$T$ candidates and estimation of the
phonon contributions to $G(T)$.

%%%%%%%%%%%%%%%%%%%%%%%%%%%%%%%%%%%%%%%%%%%%%%%%%%%%%%%%%%%%%%%%%%%%%%%%
\section{Conclusions}
\label{section.conclusions}
%%%%%%%%%%%%%%%%%%%%%%%%%%%%%%%%%%%%%%%%%%%%%%%%%%%%%%%%%%%%%%%%%%%%%%%%

The presented results illustrate that identification of ground states
depends not only on the exhaustive sampling of possible structures and
compositions but also on the accuracy of the chosen simulation
method. We have demonstrated that inclusion of the dispersive
interactions via nonlocal correlation functional within vdW-DF leads
to a good agreement with experiment for the Mg(BH$_4$)$_2$ low-$T$
ground state. The related well-characterized Ca(BH$_4$)$_2$ system
should be an interesting test case for the vdW-DF and PBE-D* methods
since standard DFT approximations seem to already reproduce the
experiment\cite{CaBH42}. Proper treatment of the nonlocal correlation
effects is expected to be most critical in higher-valent metal
borohydrides that can form weakly interacting molecular-type
complexes\cite{Lodziana}. Until such large crystalline systems can be
handled with accurate quantum Monte Carlo-level or quantum chemistry
methods, the continued improvement of DFT-based methods
\cite{progress} may provide an attractive alternative for description
of the dispersive interactions therein. In any case, it appears to be
good practice to systematically test proposed ground states with a
range of DFT flavours as this can help spot and avoid potential
artifacts of DFT approximations.

%%%%%%%%%%%%%%%%%%%%%%%%%%%%%%%%%%%%%%%%%%%%%%%%%%%%%%%%%%%%%%%%%%%%%%%%
\section{Acknowledgments}
\label{section.acknowledgments}
%%%%%%%%%%%%%%%%%%%%%%%%%%%%%%%%%%%%%%%%%%%%%%%%%%%%%%%%%%%%%%%%%%%%%%%%

A.B. and A.N.K. acknowledge the support of the EPSRC through the
DTI Technology Program EP/D062098/1 and the CAF EP/G004072/1.
Calculations were performed in the Gdansk Supercomputing Center,
on the DEAC cluster of Wake Forest University, and in the Oxford
Supercomputing Centre. A.B. thanks Bernhard Seiser for helpful
discussions.

%%%%%%%%%%%%%%%%%%%%%%%%%%%%%%%%%%%%%%%%%%%%%%%%%%%%%%%%%%%%%%%%%%%%%%%%
\appendix
\section{Benchmark tests}
\label{section.appendix}
%%%%%%%%%%%%%%%%%%%%%%%%%%%%%%%%%%%%%%%%%%%%%%%%%%%%%%%%%%%%%%%%%%%%%%%%

To estimate how technical factors, such us a choice of a
pseudopotential, may influence the accuracy of total enthalpy
calculations we performed an optimization of a set of exemplary
structures at the PBE-D* level using the projector augmented-wave
pseudopotentials (PAW)\cite{bloechl994}, as implemented in {\small
VASP}, as well as the ultrasoft pseudopotentials
(USPP)\cite{USPP}, as implemented in PWscf (see Table \ref{T2}).
In the first case, we used {\small QMPOT} software linked to
{\small VASP} to calculate dispersion contributions to energy and
forces. A crystal cell has been optimized in a series of
independent runs for a set of cell parameters. For all exemplary
structures the relative difference in cell parameters optimized
using PAW pseudopotential ({\small VASP}+{\small QMPOT}) and USPP
pseudopotential (PWscf software) was below 0.5$\%$. The difference
between relative energies calculated with PAW and USPP was found
to be below 0.0035 eV/BH$_4$. We estimate that for a given {\it
xc} functional the errors arising from other factors (convergence
criteria, choice of pseudopotential, ect.)  are rather small
allowing us to resolve polymorphs $\sim 0.005$ eV/BH$_4$ apart.
%%%%%%%%%%%%%%%%%%%%%%%%Table 2%%%%%%%%%%%%%%%%%%%%%%%%%%%%%%%%
\begin{widetext}
\begin{center}
\begin{table}[!b]
\begin{tabular}{c|ccccccccccc}
\hline\hline
              & $\alpha$ &     2    &     3    &     4    &     5    &     6    &     8    &    10    &     11    &    12    &    22     \\ \hline
  PAW         &  0.0000  &  0.0065  &  0.0029  &  0.0022  &  0.0188  &  0.0670  &  0.0671  &   0.0373 &  0.0848  &  0.0965  &  0.1091   \\
  USPP        &  0.0000  &  0.0065  &  0.0041  &  0.0022  &  0.0179  &  0.0660  &  0.0658  &   0.0406 &  0.0845  &  0.0968  &  0.1066   \\
  difference  &  0.0000  &  0.0000  & -0.0012  &  0.0000  &  0.0009  &  0.0010  &  0.0013  &  -0.0033 &  0.0003  &  0.0003  &  0.0025   \\
\hline\hline
\end{tabular}
\caption{ \small Relative enthalpy [eV/BH$_4$] of selected
polymorphs calculated at the PBE-D* level with different
pseudopotentials. }\label{T2}
\end{table}
\end{center}
\end{widetext}
%%%%%%%%%%%%%%%%%%%%%%%%%%%%%%%%%%%%%%%%%%%%%%%%%%%%%%%%%%%%%%%

%%%%%%%%%%%%%%%%%%%%%%%%%%%%%%%%%%%%%%%%%%%%%%%%%%%%%%%%%%%%%%%%%%%%%%%%


\begin{thebibliography}{99}
\bibitem{supercells}
We used $1 \times 1 \times 1$ simulation cell for $\alpha$,
P3$_1$12, Pbca, Fddd and $2 \times 2 \times 2$ simulation
supercell for I4$_1$/acd, I4$_1$/amd, F222, P$\overline{3}$m1,
respectively.
\end{thebibliography}
\end{document}